\documentclass[superscriptaddress,groupedaddress,nofootnoteinbib,12pt]{article} 
\pdfoutput=1
\usepackage{color}
\usepackage[pdftex]{graphicx}  
\usepackage{dcolumn}   
\usepackage{bm}       
\usepackage{amssymb}
\usepackage{amsmath}
\usepackage{sectsty}
\usepackage{latexsym}
\usepackage{float}
\usepackage{ifthen}
\usepackage{caption,subfig}
\usepackage{enumerate}
\usepackage{url}
\usepackage{caption,subfig}
\usepackage{feynmf}	
\usepackage{jcappub}
\usepackage{amsopn}
\usepackage{float}
\usepackage{pst-pdf}
\usepackage{amsfonts}
\usepackage{multirow}
\usepackage{array}
\usepackage{booktabs}
\usepackage{rotating}

\def\clap#1{\hbox to 0pt{\hss#1\hss}}

\def\({\left(}
\def\){\right)}
\def\[{\left[}
\def\]{\right]}
\def\bea{\begin{eqnarray}}
\def\eea{\end{eqnarray}}
\def\be{\begin{equation}}
\def\ee{\end{equation}}
\def\ba{\begin{eqnarray}}
\def\ea{\end{eqnarray}}
\def\beq{\begin{eqnarray}}
\def\eeq{\end{eqnarray}}

\def\d{\mathrm{d}}
\newcommand{\cs}{c_s}

\def\clap#1{\hbox to 0pt{\hss#1\hss}}

\newcommand{\fNL}{f_{\mathrm{NL}}}
\newcommand{\SigmaP}{\Sigma}
\newcommand{\lambdaP}{\lambda}
\newcommand{\e}{\varepsilon}
\newcommand{\im}{\mathrm{i}}

\definecolor{forestgreen}{rgb}{0.133,0.545,0.133}

\renewcommand{\d}{\mathrm{d}}
\renewcommand{\vec}[1]{\bm{\mathrm{{#1}}}}

\newcommand{\para}[1]{\par\vspace{2mm}\noindent\emph{{#1}}.---}

 \def\be   {\begin{equation}}   \def\ee   {\end{equation}}

 \def\ba  {\begin{eqnarray}}   \def\ea  {\end{eqnarray}}

\renewcommand{\thesubsection}{\arabic{section}.\arabic{subsection}}

\begin{document}

\title{Why does the effective field theory of inflation work?}

\author{Nishant Agarwal,$^{a}$ Raquel H. Ribeiro,$^{b}$ and R. Holman$^{c}$}
\affiliation{$^{a}$McWilliams Center for Cosmology, Department of Physics, Carnegie Mellon University, \\ Pittsburgh, PA 15213, USA}
\affiliation{$^{b}$Department of Physics, Case Western Reserve University, \\ 10900 Euclid Ave, Cleveland, OH 44106, USA}
\affiliation{$^{c}$Department of Physics, Carnegie Mellon University, \\ Pittsburgh, PA 15213, USA}

	\emailAdd{nishanta@andrew.cmu.edu}
	\emailAdd{raquelhribeiro@case.edu}
	\emailAdd{rh4a@andrew.cmu.edu}

\abstract{The effective field theory (EFT) of inflation has become the preferred method for computing cosmological correlation functions of the curvature fluctuation, $\zeta$. It makes explicit use of the soft breaking of time diffeomorphisms by the inflationary background to organize the operators expansion in the action of the Goldstone mode $\pi$ associated with this breaking. Despite its ascendancy, there is another method for calculating $\zeta$ correlators, involving the direct calculation of the so-called Horndeski action order by order in powers of $\zeta$ and its derivatives. The question we address in this work is whether or not the $\zeta$ correlators calculated in these seemingly different ways are in fact the same. The answer is that the actions to cubic order in either set of variables do indeed give rise to the same $\zeta$ bispectra, but that to make this equivalence manifest requires a careful understanding of the non-linear transformations relating $\pi$ to $\zeta$ and how boundary terms in the actions are affected by imposing this relation. As a by product of our study we find that the calculations in the $\pi$ language can be simplified considerably in a way that allows us to use only the {\em linear} part of the $\pi-\zeta$ relation simply by changing the coefficients of some of the operators in the EFT. We also note that a proper accounting of the boundary terms will be of the greatest importance when computing the bispectrum for more general initial states than the Bunch--Davies one.}


\maketitle


\section{Introduction}
\label{sec:intro}

The microwave sky has been mapped with incredible precision by the \emph{Planck} satellite \cite{Ade:2013nlj, Ade:2013uln, Ade:2013ydc}. To make best use of the data requires a well-defined theoretical setup, in which theoretical predictions can be tested directly against observations. In practice, given a set of observables, one can place constraints on the parameters of the theory. 

This is particularly appealing in \emph{single-clock inflation}, where there is only one active field during inflation, which is also responsible for sourcing the primordial perturbations.\footnote{We will use \emph{single-field} and \emph{single-clock} interchangeably hereafter to mean the absence of isocurvature perturbations at all times during inflation.} In this class of models, we expect there to be a finite number of parameters characterizing the field interactions. These couplings can be probed by computing $n$-point functions, which in turn build observables \cite{Maldacena:2002vr,Seery:2005wm,Chen:2006nt}. There are a number of ways in which these correlators can be computed. 

\para{Goldstone language} Inspired by particle physics ideas, one of the most natural choices of framework is the effective field theory (EFT) of single-field inflation \cite{Cheung:2007st} (see also \cite{Weinberg:2008hq} for a different setup). In this picture, because the inflaton field operates as a clock, there is a preferred choice of slicing, which corresponds to a foliation of spacetime in terms of hypersurfaces of uniform field profile, $\phi$. It follows that in this gauge the only perturbations arise from fluctuations in the metric. 

In slow-roll inflation the background softly breaks time translations, which results in a Goldstone boson, $\pi$, appearing in the particle spectrum. Consequently it is possible to write down the most general action for perturbations around a quasi-de Sitter background \cite{Creminelli:2006xe,Cheung:2007st}, in which the operators are explicitly invariant under spatial diffeomorphisms and non-linearly realized Lorentz symmetry. In this language, the small deformation of equal-$\phi$ hypersurfaces is parameterized by the Goldstone boson associated with the breaking of time-translations.

The EFT approach therefore provides a systematic algorithm for writing down the lowest dimension operators compatible with the underlying symmetries of the theory, thus performing a low-energy expansion in terms of the Goldstone mode and its derivatives. As a point of principle, this is a statement that is valid at all orders in slow-roll, with a larger number of operators being relevant the higher the order in perturbation theory we are interested in \cite{Burgess:2007pt}. However, as we shall see later, the EFT framework is particularly useful in the regime of large non-linearities, when self-interactions in the $\pi$ field dominate over the coupling to gravity. This is known as the \emph{decoupling limit}, and captures all the relevant couplings to lowest-order in slow-roll \cite{Cheung:2007st,Cheung:2007sv,Baumann:2011su}. 

In this limit and for theories in which the action only depends on the background field and its first derivatives, we can write the action for the fluctuation $\pi$ in an FRW spacetime as
\bea
	S & \supseteq & \int \d^3 x\, \d t \, a^3 \  \Bigg\{ \bar{M}^4 \[\dot{\pi}^2 - \( \dfrac{c_s}{a} \)^2 (\partial_i \pi)^2 \] +  C_{\dot{\pi}^{3}} \dot{\pi}^{3} + \frac{C_{\dot{\pi}\(\partial \pi\)^{2}}}{a^{2}} \dot{\pi}\(\partial_{i} \pi\)^{2}  \Bigg\}\,.
\label{eq:actionLOpi}
\eea
Dotted quantities above are differentiated with respect to cosmic time $t$, $\bar{M}^{4} \equiv \varepsilon H^{2} \slash c_{s}^{2}$, $\varepsilon \equiv  -\dot{H}/H^{2}$ is the usual slow-roll parameter, $H = \dot{a}/a$ is the Hubble parameter, $a(t)$ is the scale factor, and $\cs(t)$ is the sound speed of propagation of fluctuations, which need not be unity since the time-dependent background breaks Lorentz symmetry. The coefficients $C_{\dot{\pi}^3}$ and $C_{\dot{\pi} (\partial\pi)^2}$ are in general functions of the background, and measure the strength of interactions within the $\pi$ sector. In this paper we are only interested in the lowest-order statistics, and so we focus on the three-point function. Thus we ignore operators of quartic and higher order in $\pi$ when writing the action \eqref{eq:actionLOpi}.

The action above only applies for lowest order derivative theories, and in particular it does not apply to models with pathological kinetic terms, which would induce operators of the form $(\partial^2 \pi)^2$.\footnote{Nevertheless, this is not a limitation of the EFT of inflation. Indeed, such operators can be included in a way that they remain compatible with spatial diffeomorphisms. However, in this paper our interest lies in ghost-free theories, for which the action \eqref{eq:actionLOpi} is sufficient at lowest-order in slow-roll.} Moreover, the action above is to be understood symbolically in the sense that it is only valid at lowest-order in the slow-roll parameters. If slow-roll is violated, then the action \eqref{eq:actionLOpi} will have to be augmented with other operators, still cubic in $\pi$, as we discuss below.

At this order in perturbation theory, $\pi$ is conserved on super-horizon scales and its correlation functions approach a constant. However, at higher order in slow-roll the constancy of $\pi$ does not hold. Even though $\pi$ measures the small fluctuations in the local clock (using the linear relation $\delta t = -\pi = -\delta\phi/\dot{\phi}$), it is only the primordial perturbation, $\zeta$, that remains constant in the late-time limit,
\bea
	\dot{\zeta} \ \simeq \ -H \dot{\pi} + \varepsilon H^2 \pi \ \xrightarrow{(k/aH)^2 \ll 1 } \ 0 \,.
\label{eq:constzeta}
\eea
This also implies that the action \eqref{eq:actionLOpi} will not capture all the relevant physics, but will have to be augmented by other operators. In particular, for ghost-free theories and at next-to-leading order in slow-roll, we need to supplement the action in eq.\ \eqref{eq:actionLOpi} with operators of the form $\pi\dot{\pi}^2$, $\pi(\partial_i\pi)^2$, and $\dot{\pi}\partial_{i} \pi \partial^{i} \partial^{-2} \dot{\pi}$,
\bea
	S & \supseteq & \int \d^3 x\, \d t \, a^3 \  \Bigg\{ \bar{M}^4 \[\dot{\pi}^2 - \( \dfrac{c_s}{a} \)^2 (\partial_i \pi)^2+3\varepsilon H^2 \pi^2 \] +  C_{\dot{\pi}^{3}} \dot{\pi}^{3} + \frac{C_{\dot{\pi}\(\partial \pi\)^{2}}}{a^{2}} \dot{\pi}\(\partial_{i} \pi\)^{2} \nonumber \\
	& & \quad \quad \quad + \, C_{\pi\dot{\pi}^{2}} \pi\dot{\pi}^{2} + \frac{C_{\pi (\partial\pi)^{2}}}{a^2} \pi (\partial_i\pi)^{2} + C_{\rm NL} \dot{\pi} \partial_{i}\pi \partial^{i} \partial^{-2} \dot{\pi} \Bigg\} \, .
\label{eq:actionNLOpi}
\eea
We will study these operators in detail in \S\ref{sec:mystery}.

\para{Primordial perturbation language} Another picture in which one can compute correlation functions is the so-called $\zeta$- or comoving-gauge, where the propagating scalar mode is carried by the metric perturbation, $h_{ij}$, such that
\bea
	h_{ij} & = & a^2(t) e^{2\zeta} \delta_{ij} \,.
\eea
In this language, the primordial perturbation $\zeta$ becomes the perturbation of the locally defined scale factor, and it relates to the clock measuring the evolution histories of different points in an initially flat hypersurface. In other words, $\zeta=\delta N$, where $\delta N$ gives the number of elapsed e-folds during a period of inflation. 

Without commitment to a slow-roll expansion, it was shown in \cite{Gao:2011qe, DeFelice:2011uc} 
(see also \cite{Burrage:2011hd, Ribeiro:2012ar}) that expanding the action to cubic order in small $\zeta$ fluctuations, it can be written as
	\begin{equation}
		\begin{split}
		S\supseteq 
		\int \d^3 x \, \d t \; a^3  & \Bigg\{
	 z \left[
		             \dot{\zeta}^2 -\left( \dfrac{c_s}{a}\right)^2 
		             (\partial_i \zeta)^2  
		      \right] +  \Lambda_1 \dot{\zeta}^3
		      + \Lambda_2 \zeta \dot{\zeta}^2
		      +	\dfrac{\Lambda_3}{a^2} \zeta (\partial_i \zeta)^2
\\
		      &
			\ \ +  \Lambda_4 \dot{\zeta} \partial_i \zeta
				\partial^i \partial^{-2}\dot{\zeta}
			+  \Lambda_5 \partial^2 \zeta
				(\partial_i \partial^{-2}\dot{\zeta})^2	
			\Bigg\} + S_{\textrm{boundary}}\ .
		\end{split}
		\label{eq:zeta-action3}
	\end{equation}
Above, $z$ is a background dependent quantity, $\Lambda_i$ are the couplings corresponding to cubic interactions, and $S_{\textrm{boundary}}$ denotes interactions defined at the boundary \cite{Burrage:2011hd}.\footnote{In this paper we are using the results of \cite{Arroja:2011yj,Burrage:2011hd}
 who concluded that the boundary action in eq.\ \eqref{eq:zeta-action3} does not contribute to the Bunch--Davies three-point function. This implies that one has already carried out the field redefinitions discussed in \cite{Burrage:2011hd} before writing eq.\ \eqref{eq:zeta-action3}. However, and as shall be clear later, boundary terms usually \emph{do} contribute to observables and therefore cannot a priori be dismissed.} This action was originally deduced by Horndeski \cite{Horndeski:1974wa} and is frequently dubbed the \emph{Horndeski action}; it was later rederived in the context of inflation models \cite{Deffayet:2011gz,Kobayashi:2011nu,Charmousis:2011bf}. The only prerequisite for obtaining this action is that the equations of motion for the perturbations are at most second order in derivatives of the field. As a result, the action \eqref{eq:zeta-action3} applies to all single-clock inflation models, including vanilla, $k$-type \cite{ArmendarizPicon:1999rj,Garriga:1999vw}, Dirac--Born--Infeld \cite{Silverstein:2003hf,Alishahiha:2004eh} 
and galileon inflation models \cite{Burrage:2010cu}, except for ghost-inflation \cite{ArkaniHamed:2003uz}.

The advantages of this approach are: (i) all correlators built from the three-body interactions described by eq.\ \eqref{eq:zeta-action3} will remain constant on super-horizon scales since $\zeta$ does and (ii) to whatever order in slow-roll we wish to consider, there will be at most \emph{five} operators describing the cubic interactions during single-clock inflation. In particular, if there are violations of slow-roll owing, for example, to features in the potential \cite{Adshead:2012xz,Adshead:2013zfa}, the action in the $\zeta$-gauge already expresses all the relevant interactions.

Our intention in this paper is to reconcile the actions \eqref{eq:actionNLOpi} and \eqref{eq:zeta-action3} at cubic order in the fluctuations (and at next-to-leading order in slow-roll) using the non-linear relation between the two different gauges, $\pi$ and $\zeta$ (see also \cite{Gleyzes:2013ooa,Bloomfield:2013efa}). In other words, our interest lies with the bispectrum predictions in both languages, which is the lowest-order statistics sensitive to the interactions during inflation; we do not discuss higher order $n$-point functions. While we demonstrate the equivalence between these two approaches for the so-called $P(X,\phi)$ models of inflation, our analysis can be generalized to characterize the EFT coefficients in terms of the Horndeski ones or vice versa. We show that, at the level of the bispectrum, one could also use only the linear part of the relation connecting $\pi$ and $\zeta$ while carefully keeping track of boundary terms. This simplified version of the usually non-linear field redefinition should enable a more transparent comparison to observations. We also investigate the effect of boundary terms in the action for non-Bunch--Davies initial states.

\para{Outline} This paper is organized as follows. In \S \ref{sec:mystery} we obtain the action describing cubic interactions in the Goldstone boson sector from the action of the primordial perturbation. We explain in \S \ref{sec:resolution} how to obtain equivalent three-point correlators from these actions involving different interaction channels. We argue that there is a simpler but indistinguishable form of the EFT action, for which boundary terms do not contribute to the bispectrum for Bunch--Davies initial states. In \S \ref{sec:discussion} we elaborate on the role of the interactions defined at the boundary for non-Bunch--Davies initial states, and we summarize our results in \S \ref{sec:summary}.

\para{Notation} We use units in which the reduced Planck mass, $M_\textrm{Pl}=(8\pi G)^{-1/2}$, is set to unity, and the metric signature is mostly plus $(-,+,+,+)$. The slow-roll parameters are defined by  $\varepsilon\equiv -\dot{H}/H^2$, $\eta\equiv \d\ln{\varepsilon}/\d N$ and $s\equiv \d\ln{\cs}/\d N$, and they satisfy $\varepsilon, |\eta|, |s| \ll 1$ during inflation. To obtain correlation functions one generally invokes a slow-roll expansion, in order to control the perturbative expansion which is usually phrased in terms of Feynman diagrams. The result with the least powers in a slow-roll parameter is dubbed lowest-order, and denoted by LO. Likewise, next-to-leading order terms are denoted by NLO, and hence forth.


\section{Horndeski vs. EFT pictures}
\label{sec:mystery}

Our starting point is the action \eqref{eq:zeta-action3}. Now, in principle, all three-body interactions, whether in the bulk or defined at the boundary, contribute to the correlation functions. As stated earlier, however, with appropriate field redefinitions it can be shown that the boundary interactions in eq.\ \eqref{eq:zeta-action3} do not contribute to the three-point correlation function for Bunch--Davies initial states, so that we can and will dismiss the contribution of $S_{\textrm{boundary}}$ to the three-point correlators  in what follows. We will discuss their role for general initial states in \S\ref{sec:discussion}, but for the purposes of the present discussion it will be simpler to postpone their analysis.
 
Moreover, to make the analysis more concrete, we shall specify the coefficients $z$ and $\Lambda_i$ for theories whose Lagrangian only depends on the field profile and its first derivatives. This class of models is called $P(X,\phi)$, where $X=-g^{\mu\nu}\partial_{\mu} \phi \partial_{\nu} \phi$. In this case we identify the interaction vertices with parameters which measure the time dependence of the background \cite{Burrage:2011hd}
\begin{equation}
		\begin{aligned}
		\Lambda_1 &
		=
		\dfrac{\varepsilon}{H \cs^4}
			\left(
				1-\cs^2-2 \dfrac{\lambdaP \cs^2}{\SigmaP}
			\right) \ ,
		&
		\Lambda_2 &
		=
		\dfrac{\varepsilon}{\cs^4}
			\left[
				-3(1-\cs^2)+\varepsilon-\eta
			\right] \ ,
		&
		&
		\\
		\Lambda_3 &
		=
		\dfrac{\varepsilon}{\cs^2}
			\left[
				(1-\cs^2) +\varepsilon+ \eta -2s
			\right] \ ,
		&
		\Lambda_4 &
		=
		\dfrac{\varepsilon^2}{2 \cs^4}(\varepsilon-4) \ ,
		\ \ \ \ \ \ \ \textrm{and}
		&
		\Lambda_5
		&
		=
		\dfrac{\varepsilon^3}{4 \cs^4} \ .
		\end{aligned}
		\label{eq:lambdai}
\end{equation}
It follows that $\dot{\zeta}^3$ contributes at LO, $\zeta \dot{\zeta}^2$ and $\zeta (\partial\zeta)^2$ contribute both at LO and NLO, while the fourth operator contributes at NLO and NNLO. The last operator is highly slow-roll suppressed in this class of models and only contributes with NNLO terms. 
	
For $P(X,\phi)$ models,\footnote{Although $z$ is a generic function of the background in any given Horndeski theory, it is possible to deduce its properties for $P(X,\phi)$ models. Switching to conformal time and defining $\d y =\cs \d \tau$ \cite{Khoury:2008wj}, one can deduce $z(y) \sim (-k y)^{\frac{2\e+3s}{1-\e-s}}$ so that the spectrum of perturbations is nearly scale-invariant, in agreement with observations \cite{Ribeiro:2012ar}.}
\begin{equation}
	z=\varepsilon/\cs^2\, ,
\label{eq:zforPXphi}
\end{equation}
while 
\begin{equation}
	\lambda\equiv X^2 \frac{\partial^2 P}{\partial X^2} +\frac{2}{3}X^3  \frac{\partial^3 P}{\partial X^3} \ \ \ \textrm{and} \ \  \ 
	\Sigma= X\frac{\partial P}{\partial X} 
 +2X^2 \frac{\partial^2 P}{\partial X^2} \, .
\end{equation}
All of these quantities depend on the background cosmology. From the quadratic action in eq.\ \eqref{eq:zeta-action3}, which we shall denote by $S^{(2)}$, we can derive the equations of motion for the primordial perturbation as follows:
\begin{equation}
	\frac{\delta {\cal L}^{(2)}}{\delta \zeta}  =  -2 \frac{a^{3}\varepsilon}{c_{s}^{2}} \[ \ddot{\zeta} + 3H\dot{\zeta} - \frac{c_{s}^{2}}{a^{2}} \ \partial^{2}\zeta + (\eta -2s) H \dot{\zeta} \] + 2 \frac{a^{3}\varepsilon}{c_{s}^{2}} \dot{\zeta}\, \delta(t - t_{\rm boundary})\, ,
\label{eq:varyzeta}
\end{equation}
where ${\cal L}^{(2)}$ is the corresponding Lagrangian density. Above, we have ignored spatial boundary terms, since these terms are proportional to the total momentum, which vanishes because of momentum conservation. The last term does not contribute to the dynamics\footnote{The reader might be worried that dismissing a quadratic boundary term might result in losing information about initial conditions. However, any boundary contribution can be absorbed into a renormalization of the initial state by choosing appropriate Bogoliubov coefficients for the elementary wavefunctions of perturbations \cite{Agarwal:2012mq} (see also \cite{Holman:2007na}). The future boundary term, which appears to violate causality, on the other hand, will not contribute to the generating functional, since by construction it will cancel between the plus and minus branches of a closed-time path contour, in the language of the in--in formalism 
\cite{Schwinger:1960qe,Mahanthappa:1962ex,Bakshi:1962dv,Kadanoff:1962,Bakshi:1963bn,Keldysh:1964ud,Jordan:1986ug,Calzetta:1986ey}.}, and on-shell it suffices to use
\begin{equation}
	\frac{\delta {\cal L}^{(2)}}{\delta \zeta}  =0=  
   \ddot{\zeta} + 3H\dot{\zeta} - 
  \frac{c_{s}^{2}}{a^{2}}\ \partial^{2}\zeta + 
  (\eta -2s) H \dot{\zeta} \, .
\label{eq:varyzetafinal}
\end{equation}
The variation of the quadratic action above allows us to solve for the \textit{free} Green's function, which is then used to calculate any $n$-point correlation function, treating cubic and higher order terms in the action as interactions.

\para{Non-linear gauge transformation} The primordial perturbation, $\zeta$, and the Goldstone boson, $\pi$, can be related via the non-linear gauge field transformation:
\begin{equation}
	\zeta=-H \pi \ +f \, .
\label{eq:gaugetransf}
\end{equation}
To be more precise, it was shown in \cite{Maldacena:2002vr,Cheung:2007sv} that to quadratic order in $\pi$,
\begin{equation}
	f = - \frac{1}{2}\varepsilon H^{2}\pi^{2} +H\pi\dot{\pi} +\frac{1}{4 a^{2}} \left[ -(\partial_i\pi)^{2} + \partial^{-2}\partial_{i}\partial_{j}(\partial_{i}\pi \partial_{j}\pi) \right]\, .
\label{eq:def-f}
\end{equation}
However, we will see that only the first two terms will be important as far as the super-horizon limit of the three-point function is concerned.

\subsection{Change of gauge}

While the two parameterizations of fluctuations about the inflating background can be related via a gauge transformation, it is only $\zeta$ that remains constant outside the horizon, when spatial gradients can be safely neglected. This means that it is the statistics of $\zeta$, rather than $\pi$, that provide the dictionary between primordial inflationary physics and cosmic microwave background (CMB) and large scale structure (LSS) observations. Therefore, despite being able to write down a simple effective action in $\pi$ for modes deep inside the horizon, we need to convert $\pi$ correlators to $\zeta$ ones to evolve them outside the horizon.

\para{Quadratic action} Applying the transformation \eqref{eq:gaugetransf} to the quadratic action in eq.\ \eqref{eq:zeta-action3}, and with the identification in eq.\ \eqref{eq:zforPXphi}, we find
\begin{equation}
		\begin{aligned}
		S^{(2)}[\zeta]  \longrightarrow  &
\int{\d^3x \, \d t} \ \dfrac{a^3 \varepsilon}{\cs^2} \,H^2 
                \[
                \dot{\pi}^2 -\left(\dfrac{\cs}{a}\right)^2 (\partial_i \pi)^2
                +3\varepsilon H ^2\pi^2 \] \\
                &+\int{\d^3x \, \d t}\  f \, \left . \frac{\delta {\cal L}^{(2)}}{\delta \zeta} \right |_{\zeta = -H\pi}+
                \int_{\partial}{\d^3x } \ \dfrac{a^3 \varepsilon}{\cs^2} \(
                2f\dot{\zeta} - \varepsilon H^2\pi^2
                \) \, .
        \end{aligned}
        \label{eq:quadtransfzeta}
\end{equation}
In the first line we observe that the Goldstone boson has acquired a mass at NLO in slow-roll, which is consistent with the fact that the mixing with gravity makes $\pi$ slightly massive. The first term of the second line will not contribute to the three-point $\zeta$ correlation function, since it is a redundant operator \cite{Weinberg:1995mt}.\footnote{Note, however, that this will not be true in general for higher $n$-point functions. It is also in this sense that the conclusions of \cite{Arroja:2011yj,Burrage:2011hd} are unchanged.} Likewise, the last quadratic term does not contribute to the dynamics and can be dismissed. However, the cubic term at the boundary will in general contribute to the three-point function, and therefore must be kept \cite{Burrage:2011hd}. 

\para{Cubic action} We now analyze the effect of the gauge transformation \eqref{eq:gaugetransf} on the cubic action for primordial perturbations. For this we only require the linear piece of the field transformation in eq.\ \eqref{eq:gaugetransf}, since the non-linear contributions will only be relevant for higher $n$-point functions. Also, to make the calculation more explicit we work at NLO in slow-roll.

Before applying the gauge transformation, we notice that in the EFT approach there is a different operator from the one arising in the $\zeta$-gauge, namely $\dot{\pi} (\partial_i \pi)^2$. This is not surprising since the action itself can be made of different operators, provided the physical correlators stay invariant \cite{Arroja:2011yj,Rigopoulos:2011eq,RenauxPetel:2011sb,Ribeiro:2011ax,Battefeld:2011ut,Byun:2013jba}. With this in mind, we start by considering an operator of the form $\dot{\zeta} (\partial_i \zeta)^2$. Upon spatial and time integrations by parts we obtain
\begin{equation}
	\int \d^3 x\, \d t \ \tilde{g} \, \dot{\zeta} (\partial_i \zeta)^2 =
-\int_{\partial}{\d^3x}\ \tilde{g} \, \zeta (\partial_i \zeta)^2
+ \int{\d^3 x \, \d t} \[ 
\frac{\d}{\d t} (\tilde{g}) \, \zeta (\partial_i \zeta)^2
- 2\tilde{g} \, \zeta \dot{\zeta} \partial^2\zeta  \] \, ,
\end{equation}
where $\tilde{g}$ is the interaction vertex associated with such an operator. Using the on-shell equation of motion \eqref{eq:varyzetafinal} and setting
\bea
	\tilde{g} & = & -(1-\cs^2) \, \dfrac{a\e}{H\cs^2}\, ,
\eea
we find
\begin{equation}
		\begin{aligned}
		\int \d^3 x\, \d t \ \tilde{g}\, \dot{\zeta} (\partial_i \zeta)^2  
		\, = \, &
                \int{\d^3x \, \d t} \ \tilde{g} \ 
                \Bigg\{
                   \left(\dfrac{a}{\cs}\right)^2  \dot{\zeta}^3
                   + H  \left(\dfrac{a}{\cs}\right)^2 
                   \( -\frac{2s \cs^2}{1-\cs^2}-3-\eta+\e \)
                   \zeta \dot{\zeta}^2 \\
                   & \hspace*{2.8cm}
                 + \ \[ -\frac{2sH \cs^2}{1-\cs^2}+H 
                 (1+\eta+\e-2s) \] \zeta (\partial_i \zeta)^2
                \Bigg\} \\
            & - \int_\partial \d^3 x \ \dfrac{\tilde{g}a^2}{\cs^2} \[
            \zeta \dot{\zeta}^2 +
            \left(\dfrac{\cs}{a}\right)^2 \zeta (\partial_i \zeta)^2
            \] \, .    
        \end{aligned}    \label{eq:integ-operator}
\end{equation}
We conclude that the correlation built out from the three-body interaction $\dot{\zeta} (\partial_i \zeta)^2$ is equivalent to a linear combination of operators both in the bulk and at the boundary. Notice that all the operators on the right-hand side of the equation above are already included in the original action for $\zeta$, so we can simply add and subtract eq.\ \eqref{eq:integ-operator} from the action in eq.\ \eqref{eq:zeta-action3}. Further, we also need to account for the cubic boundary term obtained in eq.\ \eqref{eq:quadtransfzeta}, which was generated by applying the field redefinition to the quadratic action. Moreover, we formally keep the boundary terms originally present in the cubic action and represented by $S_{\textrm{boundary}}$. The cubic action can then be written as
\begin{equation}
		\begin{aligned}
		S^{(3)} [\zeta]
		\, = \, &
                \int{\d^3x \, \d t} \ 
                \bigg[
                -\frac{2\lambda}{\Sigma} \frac{\e a^3}{H\cs^2} \dot{\zeta}^3
        -\tilde{g} \,\dot{\zeta} (\partial_i\zeta)^2 \\      
	& \hspace*{0.3cm} + \frac{a^3\e}{\cs^2} (2s+\e-\eta)
                   \zeta \dot{\zeta}^2
                   +a\e(\e+\eta)\zeta (\partial_i \zeta)^2            
                 -2\dfrac{a^3\varepsilon^2}{\cs^4} 
               \dot{\zeta} \partial_i \zeta
				\partial^i \partial^{-2}\dot{\zeta} \bigg] \\
            & - \int_\partial \d^3 x \ \dfrac{\tilde{g}a^2}{\cs^2} \[
            \zeta \dot{\zeta}^2 +
            \left(\dfrac{\cs}{a}\right)^2 \zeta (\partial_i \zeta)^2
            \] + 2\int_{\partial}{\d^3x } \ \dfrac{a^3 \varepsilon}{\cs^2}\ 
                f\dot{\zeta} 
               + S_{\textrm{boundary}} \, .
        \end{aligned}
\end{equation}
Now, performing only the required linear part of the field redefinition in eq.\ \eqref{eq:gaugetransf} on the \emph{bulk} operators, it follows that
\bea
		S^{(3)} [\zeta] & = &
                \int{\d^3x \, \d t} \, \dfrac{a^3H^3}{\cs^2}  
                \bigg[ \frac{2\lambda}{\Sigma} \frac{\e}{H} 
                  \dot{\pi}^3
	+ \frac{\cs^2}{a^3}\tilde{g} \,\dot{\pi} (\partial_i\pi)^2 \nonumber \\
		 & & \quad 
                   + \e 
                   \Big(
                   -\frac{6\lambda}{\Sigma}\varepsilon -2s -\e+\eta
                   \Big)
                   \pi \dot{\pi}^2
                   + \frac{\e}{a^2}
                   \left(-2\e \cs^2+\e-\eta\cs^2
                    \right)  
                   \pi (\partial_i \pi)^2  
                 + 2\dfrac{\varepsilon^2}{\cs^2}  
               \dot{\pi} \partial_i \pi
				\partial^i \partial^{-2}\dot{\pi} \bigg] \nonumber \\
            & & - \int_\partial \d^3 x \ \dfrac{\tilde{g}a^2}{\cs^2} \[
            \zeta \dot{\zeta}^2 + 
            \left(\dfrac{\cs}{a}\right)^2 \zeta (\partial_i \zeta)^2
            \] + 2\int_{\partial}{\d^3x } \ \dfrac{a^3 \varepsilon}{\cs^2}\ 
                f\dot{\zeta} +S_{\textrm{boundary}} \, .
\eea

Why didn't we apply the field redefinition to the boundary terms as well? For one thing, the first integral on the boundary actually does not contribute to the three-point correlator \emph{if} the initial state is Bunch--Davies
(this can be checked by explicit computation). The $\zeta(\partial_i\zeta)^2$ term does not contribute simply because any boundary operators with no time derivatives only produce a phase which cancels between the plus and minus contours in the in--in picture. The $\zeta\dot{\zeta}^2$ term, on the other hand, can be canceled by performing a field redefinition such that the bispectrum remains invariant, as was shown in \cite{Arroja:2011yj,Burrage:2011hd}. Also, the boundary terms denoted as $S_{\rm boundary}$ do not contribute to the three-point correlator for Bunch--Davies initial states, as mentioned earlier. Finally, the second integral on the boundary, which has a cubic interaction of the form $f\dot{\zeta}$ \emph{will} contribute to the bispectrum even for Bunch--Davies initial states, and so we will keep it in the analysis that 
follows.\footnote{Choosing an initial state different from Bunch--Davies will have an effect at the early time boundary; we will come back to this point in \S\ref{sec:discussion}. On the other hand, whether an interaction operator contributes to the bispectrum at the late time boundary should be independent of the choice of initial state, and rather only be a function of the derivative structure of the operator itself---we refer the reader to refs. \cite{Arroja:2011yj,Burrage:2011hd}, where this point was thoroughly investigated.}

Hence, for Bunch--Davies initial states, it suffices to consider the following action at the level of the generating functional for the three-point correlations:
\bea
		S^{(3)}_{\rm BD} [\zeta] & \supseteq &
                \int{\d^3x \, \d t} \, \dfrac{a^3H^3}{\cs^2}  
                \bigg[ \frac{2\lambda}{\Sigma} \frac{\e}{H} 
                  \dot{\pi}^3
		- (1-c_s^2) \frac{\varepsilon}{a^2 H} \,\dot{\pi} (\partial_i\pi)^2 \nonumber \\
		 & & \quad
                   + \, \e 
                   \Big(
                   -\frac{6\lambda}{\Sigma}\varepsilon -2s -\e+\eta
                   \Big)
                   \pi \dot{\pi}^2 + \frac{\e}{a^2}
                   \left(-2\e \cs^2+\e-\eta\cs^2
                    \right)  
                   \pi (\partial_i \pi)^2 
                 + 2\dfrac{\varepsilon^2}{\cs^2}  
               \dot{\pi} \partial_i \pi
				\partial^i \partial^{-2}\dot{\pi} \bigg] \nonumber \\
            & & + \, 2\int_{\partial}{\d^3x } \ \dfrac{a^3 \varepsilon}{\cs^2}\ 
                f\dot{\zeta} \, .
\label{eq:S3pibd}
\eea

To reach this conclusion we have performed the full non-linear field redefinition in eq.\ \eqref{eq:gaugetransf}. If we wish to predict physical observables from correlations, then we need to compute the $\zeta$ correlators in terms of the $\pi$ correlators. Since we have performed a non-linear transformation between the $\zeta$ and $\pi$ gauges, $\langle \zeta \zeta \zeta \rangle$ is different from $-H^3 \langle \pi \pi \pi \rangle$. 
On the other hand, if we were to perform only the \emph{linear} transformation $\zeta = -H\pi$ throughout, then we would precisely obtain $\langle \zeta \zeta \zeta \rangle = -H^3 \langle \pi \pi \pi \rangle$, and additionally there would be no boundary term in the above action.\footnote{Notice that the linearized field redefinition we refer to is only linear in $\pi$, but is exact in slow-roll (i.e., it need not be supplemented with slow-roll corrections).} We will demonstrate explicitly how these approaches can be reconciled in \S \ref{sec:resolution}.

\subsection{Contact with the EFT action}

How does eq.\ \eqref{eq:S3pibd} relate to the EFT action for $\pi$? The form of the operators 
matches, but the coefficients \emph{do not}. Indeed, relating the background quantity $\lambda/\Sigma$ to the parameters in the EFT, we learn that
for $P(X,\phi)$ models \cite{Cheung:2007st,Cheung:2007sv}
\bea
	2\, \frac{\lambda}{\Sigma} & = & (1-\cs^2) \left(1+\frac{2}{3} \frac{M_3^4}{M_2^4} \right) \, ,
\eea
and using this in eq.\ \eqref{eq:S3pibd} gives
\bea
	S^{(3)}_{\rm BD}[\pi]  & = & \int \d^{3}x \, \d t \, a^{3} \, \bigg\{ C_{\dot{\pi}^{3}} \dot{\pi}^{3} + \frac{C_{\dot{\pi}(\partial\pi)^{2}}}{a^{2}} \dot{\pi}(\partial_i\pi)^{2} \nonumber \\
	& & \quad \quad \quad + \, C_{\pi\dot{\pi}^{2}} \pi\dot{\pi}^{2} + \frac{C_{\pi (\partial\pi)^{2}}}{a^2} \pi (\partial_i\pi)^{2} + C_{\rm NL} \dot{\pi} \partial_{i}\pi \partial^{i} \partial^{-2} \dot{\pi} \bigg\} \nonumber \\
	 & & + \, 2\int \d^{3}x \, \d t \, \frac{a^{3}\epsilon^{2} H^{3}}{c_{s}^{2}} \pi \left[ \dot{\pi}^{2} - \frac{c_{s}^{2}}{a^{2}} (\partial_i\pi)^{2} \right] + 2\int_{\partial}{\d^3x } \ \dfrac{a^3 \varepsilon}{\cs^2} \ f\dot{\zeta} \, .
\label{eq:cubicactionpi}
\eea
The first two lines are precisely the EFT action with interaction coefficients given by \cite{Cheung:2007sv}
\begin{equation}
		\begin{aligned}
		C_{\dot{\pi}^{3}} & =  \bar{M}^{4} (1-c_{s}^{2}) 
		\left( 1 + \frac{2}{3} \frac{M_3^{4}}{M_2^{4}} \right) 
		\ ,
		&
		C_{\dot{\pi}\left(\partial\pi\right)^{2}} &
		=-
		\bar{M}^{4} \left(1 - \cs^{2}\right) \ , & 
		\\
		C_{\pi\dot{\pi}^{2}} &
		=
		\bar{M}^{4}H \left[ -6\e + \eta - 2s + 3\e \cs^{2} 
		- 2\e \frac{M_3^{4}}{M_2^{4}} \left(1-\cs^{2}\right) \right] \ ,
		&
		C_{\pi\left(\partial\pi\right)^{2}} &
		=
		\bar{M}^{4}H \left( \e - \eta \cs^{2} \right) \ , \\
		C_{\rm NL}
		 &=
		\bar{M}^{4}H \left( \frac{2\e}{\cs^{2}} \right) \ , & &\ \ 
		\end{aligned}
		\label{eq:Ci}
\end{equation}
where again $\bar{M}^4 \equiv \e  H^2/\cs^2$. Additionally, we find two bulk operators and a boundary term in the second line, all of which contribute to the correlators even for Bunch--Davies initial states, and are usually not mentioned in the literature. Let us first focus on the extra operators in the bulk. At NLO in slow-roll, we integrate by parts to obtain
\bea
	S^{(3)}_{\rm extra}[\pi] & = & 2\int \d^{3}x \, \d t \, \frac{a^3\e^{2} H^3}{\cs^2} \pi 
\left[ \dot{\pi}^2 - \frac{\cs^2}{a^2} (\partial\pi)^2 \right] \nonumber \\
	& = & - \int \d^3 x \, \d t \, \frac{a^3\e^2 H^{3}}{\cs^2} \pi^2 \left[ \ddot{\pi} + 3H\dot{\pi} - \frac{\cs^2}{a^2} \partial^{2}\pi \right]  + \ \int_{\partial} \d^{3}x \, \frac{a^3\e^2 H^{3}}{\cs^2} \pi^{2} \dot{\pi} \, .
\label{eq:cubicextraa}
\eea
One can explicitly show that the bispectrum resulting from the bulk term in the last line is zero. Alternatively, one can invoke Weinberg's procedure \cite{Weinberg:1995mt} for dealing with such a redundant operator, as before.
As a result, we finally obtain
\bea
	S^{(3)}_{\rm extra}[\pi]\supseteq  \int_{\partial} \d^3 x \, \frac{a^3\e^2 H^3}{\cs^2} \pi^2 \dot{\pi} \, .
\label{eq:cubicextrab}
\eea

Consequently, the extra bulk operators are actually equivalent to a single-derivative operator acting on the time boundary, which contributes at NLO.\footnote{This can also be checked explicitly by computing the bispectrum.} Although we have made some progress by pushing the contribution to the boundary, this term is nonetheless dangerous because it cannot be gauged away and \emph{will} contribute to the bispectrum of perturbations \cite{Burrage:2011hd}, irrespective of the choice of initial conditions. How then can we obtain identical observables in the Horndeski and EFT approaches?


\section{Construction of observables}
\label{sec:resolution}

Ultimately we will be interested in correlators of the primordial fluctuation, $\zeta$, either built from the Horndeski action, or from the EFT language. Regardless of the choice of action, the generating functional of correlations in $\zeta$ should agree as it should describe the same physical effects, even if the two actions involve different operators and fields.

What we have found in the last section, however, is a different issue. Starting from the action of perturbations, we obtain the standard EFT action but with two extra boundary operators,
\bea
	\int_\partial \d^3 x \, \frac{a^3 \varepsilon^2 H^3}{c_s^2} \pi^2 \dot{\pi} \ \ \ \ \textrm{and} \ \ \ 2\int_\partial \d^3 x \, \frac{a^3 \varepsilon}{c_s^2} f \dot{\zeta} \, .
\label{eq:extrabdyterms}
\eea
For Bunch--Davies initial states, only the late time boundary in each of these terms will contribute. The contribution coming from the early time boundary can be set to zero using the standard $i\epsilon$ prescription. Furthermore, since $\zeta$ is constant at late times, we can set the spatial derivatives of $\pi$ to zero in $f$, and $\dot{\pi} = \varepsilon H\pi$. Therefore, outside the horizon we have $f = \frac{1}{2} \varepsilon H^2 \pi^2$. At the appropriate order in slow-roll, $\dot{\zeta}$ is simply $-H\dot{\pi}$, so that the boundary terms written in eq.\ \eqref{eq:extrabdyterms} exactly cancel.

In this sense, starting with the Horndeski action for $P(X,\phi)$ models and assuming Bunch--Davies initial conditions, we can obtain the action for the Goldstone mode $\pi$ using the full non-linear field redefinition in eq.\ \eqref{eq:gaugetransf}. The three-point function for $\zeta$ can then be related to to the three-point function for $\pi$ in the usual way: $\langle \zeta \zeta \zeta \rangle = -H^3 \langle \pi \pi \pi \rangle$ plus a piece that comes from the non-linear part of the transformation, which acts as a convolution in momentum space \cite{Cheung:2007sv}.

Let us now consider what would happen if we were to perform only the \emph{linear} transformation $\zeta = -H\pi$ throughout. This is equivalent to setting $f$ to zero everywhere. In that case, we will still obtain the EFT action, but without the second boundary term written in eq.\ \eqref{eq:extrabdyterms}. The $\pi^2\dot{\pi}$ boundary term will therefore not cancel. Further, at the level of the three-point function, we will now have exactly $\langle \zeta \zeta \zeta \rangle = -H^3 \langle \pi \pi \pi \rangle$. So one may guess that the contribution of the convolution to the three-point correlator in $\zeta$ when we performed the full non-linear field redefinition should exactly equal the contribution of the first boundary term in eq.\ \eqref{eq:extrabdyterms} to the three-point function. As we show below, this guess is indeed correct.

Evaluating the impact of the $\pi^2\dot{\pi}$ boundary operator with the special choice of starting in the standard vacuum, we find that it contributes to the bispectrum of perturbations as follows
\bea
	\langle \zeta(\vec{k}_1) \zeta(\vec{k}_2) \zeta(\vec{k}_3) \rangle & \supseteq & (2\pi)^3 \delta^{(3)} \( \sum {\vec k}_i \) \frac{H^4}{16\e \cs^2} \frac{\sum_i k_i^3}{\prod_i k_i^3} \, = \, (2\pi)^3 \delta^{(3)} \( \sum {\vec k}_i \) \varepsilon \sum_{i\neq j} P(k_i) P(k_j) \, , \nonumber \\
\label{eq:extrabisp}
\eea
where $P(k_i)$ is the primordial power spectrum in $P(X,\phi)$ theories, $P(k_i) = \frac{H^2}{4\e\cs k_i^3}$. The result in eq.\ \eqref{eq:extrabisp} is non-zero, as anticipated. In fact it corresponds to a local contribution to the bispectrum, and therefore its contribution is largest in the so-called squeezed limit, when one of the momenta becomes much smaller than the other two.

Now we will review the contribution to the bispectrum from the non-linear part of the field redefinition that is usually invoked, following the discussion in \cite{Cheung:2007sv}. Since we are interested in the correlation function at late times, we can again set $f = \frac{1}{2} \varepsilon H^2 \pi^2$. Then in Fourier space we have
\begin{equation}
	\zeta (\vec{k}) = -H \pi (\vec{k}) + \frac{1}{2} \e H^2 \int{\frac{\d^3 \vec{q}}{(2\pi)^3}} \pi(\vec{q}) \pi(\vec{k}-\vec{q}) \, ,
\label{eq:convolution}
\end{equation}
where the last term is a convolution in momentum space. Using eq.\ \eqref{eq:convolution} to compute the three-point function gives
\bea
	\langle \zeta(\vec{k}_1) \zeta(\vec{k}_2) \zeta(\vec{k}_3) \rangle & = & - H^3 \langle \pi(\vec{k}_1) \pi(\vec{k}_2) \pi(\vec{k}_3) \rangle \nonumber \\
	& & \quad + \ \frac{\e H^4}{2} \int \frac{\d^3 \vec{q}}{(2\pi)^3} \langle \pi(\vec{k}_1) \pi(\vec{k}_2) \pi(\vec{q}) \pi(\vec{k}_3-\vec{q}) \rangle + {\rm c.p.} \nonumber\\
	& = & - H^3 \langle \pi(\vec{k}_1) \pi(\vec{k}_2) \pi(\vec{k}_3) \rangle \nonumber \\
	& & \quad + \ \e H^4 \int \frac{\d^3 \vec{q}}{(2\pi)^3} \langle \pi(\vec{k}_1) \pi(\vec{q}) \rangle
\langle \pi(\vec{k}_2) \pi(\vec{k}_3-\vec{q}) \rangle + {\rm c.p.} \, .
\label{eq:convfinal}
\eea
In the last line we have performed Wick contractions leading only to connected diagrams, and `c.p.' encodes the cyclic permutations over momenta.

Using the quadratic EFT action written in the first line of eq.\ \eqref{eq:quadtransfzeta} to construct the generating functional in $\pi$, one can easily obtain that $P_{\pi}(k_i) \equiv \langle \pi(\vec{k}_i) \pi(\vec{k}_j) \rangle = P(k_i)/H^2$, and hence
\bea
	\langle \zeta(\vec{k}_1) \zeta(\vec{k}_2) \zeta(\vec{k}_3) \rangle & = & - H^3 \langle \pi(\vec{k}_1) \pi(\vec{k}_2) \pi(\vec{k}_3) \rangle + (2\pi)^3 \delta^{(3)} \( \sum {\vec k}_i \) \varepsilon \sum_{i\neq j} P(k_i) P(k_j) \, .
\eea
The contribution from the convolution is therefore precisely equal to the quantity in eq.\ \eqref{eq:extrabisp}  that we obtained from the boundary term that survives on carrying out only a linear gauge transformation. 

We conclude that the effect of the non-linear part of the field transformation \eqref{eq:gaugetransf} is to absorb the boundary term generated in writing the Horndeski action for $P(X,\phi)$ theories in the EFT form with the usual coefficients, when carrying out only the \emph{linear} field transformation. Ultimately we know that the observables constructed out of correlation functions of $\zeta$ have to match in both approaches. Seeing how this explicitly works out for the three-point function clarifies what assumptions are made in showing this equivalence. In particular, one important assumption we made was that the initial state is the Bunch--Davies vacuum. Nevertheless, boundary terms can, in general, have non-trivial effects when the initial time boundary is not taken to be at $\tau_0 \rightarrow -\infty$. We discuss this issue further in the following section.


\section{Discussion}
\label{sec:discussion}

The results from the previous sections are encouraging for a number of reasons. First, we have deduced a simpler form for the EFT action from the Horndeski action by performing a linear gauge transformation from $\zeta$ to $\pi$. The result of this is the bulk action in eq.\ \eqref{eq:S3pibd}, which contains all operators at NLO compatible with the symmetries during slow-roll inflation. This bulk action differs from the standard EFT action by the extra operators considered in eq.\ \eqref{eq:cubicextrab}. On the other hand, had we performed the non-linear field redefinition there would also be an additional operator defined at the boundary, which precisely cancels the contribution from these extra operators for Bunch--Davies initial states. In addition, we also have boundary operators carried over from the $\zeta$-gauge, which again do not contribute for Bunch--Davies initial states. When do these boundary terms become important?

\para{General initial states} Part of the theoretical setup in obtaining correlation functions is choosing an appropriate initial state. The standard, simplest prescription uses the Bunch--Davies vacuum \cite{Bunch:1978yq,Birrell:1982ix}, and considerable effort has been lent to understanding what other appropriate choices can be made \cite{Kaloper:2002cs,Collins:2003mj,Greene:2005aj,ArmendarizPicon:2006pd}. If the standard Bunch--Davies initial state is defined by the LO wavefunction for the primordial perturbation
\begin{equation}
	u_k(\tau) = -\frac{\im}{2} \frac{H}{\sqrt{\e/\cs^2}} \frac{1}{(\cs k)^{3/2}} 
(1+\im \cs k \tau)\, e^{-\im \cs k \tau} \, ,
\end{equation}
then an arbitrary initial state can be described by a generalized wavefunction
\begin{equation}
	w_k(\tau) = \alpha_k u_k (\tau)+\beta_k u_k^{\star} (\tau) \, .
\end{equation}
Above, $\alpha_k$ and $\beta_k$ are the Bogoliubov coefficients, which satisfy the Wronskian condition $|\alpha_k|^2-|\beta_k|^2=1$. This \emph{new} vacuum has a different particle spectrum, with number density in $k$-space given by $|\beta_k|^2$.

Importantly, if inflationary correlators depend on the choice of initial conditions, we might be able to use observational data as a means to constrain such modifications \cite{Contaldi:1999jr, Chen:2006nt, Holman:2007na, Collins:2009pf, Meerburg:2009ys, Meerburg:2009fi, Agullo:2010ws, Ashoorioon:2010xg, Ganc:2011dy, Agullo:2011xv, Chialva:2011hc, Ganc:2012ae, Agullo:2012cs, Agarwal:2012mq, Agarwal:2013bya, Flauger:2013hra, Aravind:2013lra, Ashoorioon:2013eia, Brahma:2013rua, Bahrami:2013isa, Agarwal:2013qta}. In particular, one of the key insights is that statistics with a non-Bunch--Davies choice of initial state have a characteristic signature. The amplitude of the three-point function in this case, usually called $\fNL$, is enhanced for flattened ($k_3 = k_1 + k_2$) and squeezed ($k_3 \ll k_1 \approx k_2$) configurations of momenta. Therefore if this signal is observed we may be able to use it to obtain information on the fossilized initial state of the early Universe.

\para{Statistics dependence on the initial state} How are inflationary correlators computed for a general initial state? Correlation functions are calculated using the in--in prescription which is appropriate to evaluate expectation values at some fixed time. The computation involves performing a path integral over the possible field configurations, such that the correlation function of a product of $n$ copies of the primordial perturbation becomes: 
\begin{equation}
	\left<\zeta_1(\vec{k}_1)\cdots \zeta_n(\vec{k}_n) \right> (\tau)=
\int{[\mathcal{D} \zeta^{+}]\, [\mathcal{D} \zeta^{-}]} \
\zeta_1(\tau,\vec{k}_1) \cdots \zeta_n(\tau,\vec{k}_n)  e^{\im (S[\zeta^+]-S[\zeta^-])}\, 
\delta[\zeta^+ (\tau_\star)-\zeta^- (\tau_\star)]\ .
\label{eq:correlatorinin}
\end{equation}
The forward path integral is labeled by the field configuration $\zeta^+$ whilst the backwards one is labeled by $\zeta^-$. In practice this can be done by performing a time integral from an initial time (at which the vacuum of the theory is defined) to some time, say $\tau_\star$, later than the time of interest for the expectation value. To this we add the path integral performed backwards, which returns to the vacuum at some initial time. The Dirac distribution constrains the fields $\zeta^+$ and $\zeta^-$ to agree at the final time $\tau_\star$. In principle a correlator such as \eqref{eq:correlatorinin} will be a function of time. However, since we are interested in the statistics of the primordial curvature perturbation on super-horizon scales, this object is guaranteed to be time independent in this regime. 

For Bunch--Davies initial conditions there is a clear prescription, first employed in the context of non-Gaussianities by Maldacena \cite{Maldacena:2002vr}. The choice of integration contour is such that this integral converges at very early times when we take the limit $\tau_0 \rightarrow -\infty$, corresponding to matching to the standard choice of vacuum. Likewise, for generalized states the choice of contour needs to reflect the initial state at early times. This is achieved by setting the lower integration limit to $\tau_0$---in the EFT picture $\tau_0$ is interpreted as the earliest time at which the theory provides a reliable description of inflationary cosmology \cite{Agarwal:2012mq}.\footnote{In the path integral formulation the choice of initial state can be expressed in terms of an appropriate initial density matrix of states $\rho$ as follows \cite{Agarwal:2012mq} 
\begin{equation*}
	\left<\zeta_1(\vec{k}_1)\cdots \zeta_n(\vec{k}_n) \right> (\tau)=
\int{[\mathcal{D} \zeta^{+}]\, [\mathcal{D} \zeta^{-}]} \
\zeta_1(\tau,\vec{k}_1) \cdots \zeta_n(\tau,\vec{k}_n) \ \rho[\zeta^+,\zeta^-;\tau_0] \ 
e^{\im (S[\zeta^+]-S[\zeta^-])}\, 
\delta[\zeta^+ (\tau_\star)-\zeta^- (\tau_\star)]\ .
\end{equation*}
}

Ultimately this has a profound impact on the computation of the bispectrum of perturbations. The arguments given in \cite{Arroja:2011yj,Burrage:2011hd} for discarding the boundary terms in the bispectrum no longer apply since they relied on the integration contour being stretched to very early times $\tau_0 \rightarrow -\infty$, thereby choosing the interacting vacuum of the theory to match Bunch--Davies. Consequently, for a generic excited initial state, any boundary terms \emph{will}, in principle, contribute to the statistics of the primordial fluctuations. In particular, the boundary terms present in eq.\ \eqref{eq:zeta-action3}, which we collectively dubbed $S_{\textrm{boundary}}$, need to be taken into account. Similarly, it might be important to include boundary operators in the EFT action, as we discuss below.

To explicitly check the relevance of boundary terms for generic initial states, one might wish to compute the bispectrum in terms of the primordial perturbation, $\zeta$, at LO in slow-roll using both the Horndeski and the EFT actions. For the sake of this comparison, let us ignore the boundary terms $S_{\textrm{boundary}}$, and only look at the transformation of the remaining terms from the $\zeta$-gauge to the $\pi$-gauge at LO in slow-roll. Computing the bispectrum for a generic initial state is now straightforward using the Horndeski action. The calculation using the EFT action is, however, a bit more subtle. First, we need to understand how the initial state transforms under the change of gauge.

Starting with a general Gaussian initial state (i.e. a Gaussian initial density matrix, in the language of \cite{Agarwal:2012mq}), a non-linear field redefinition will turn a Gaussian $\zeta$-gauge initial action into a non-Gaussian $\pi$-gauge action. Since we can choose any coefficients for operators in the $\zeta$ initial state, and therefore for resulting non-Gaussian operators in the $\pi$ initial state, it appears that predictions in the $\zeta$- and $\pi$-gauges have no reason to match. We thus adopt a simple prescription---we use only the {\it linear} field redefinition $\zeta = -H\pi$, which we have already shown to work for a Bunch--Davies initial state. With this prescription the $\zeta$- and $\pi$-gauge Green's functions are simply related by a factor of $H^2$. With this understanding we can calculate the bispectrum for a general initial state in the $\pi$ gauge.

For non-Bunch--Davies initial states we then find that predictions from both approaches only agree if we take the boundary terms generated from integrating by parts (those in eq.\ \eqref{eq:integ-operator}) into account. Note that while these boundary terms do not contribute for Bunch--Davies initial states, it is imperative to account for them to match the non-Bunch--Davies bispectra. One may also interpret these boundary terms as being part of the initial state in the $\pi$-gauge, in which case the $\pi$-gauge action need not be supplemented with additional boundary terms. As stated above, it does not appear though that predictions in both gauges will simply match by performing a non-linear gauge transformation at the level of the initial state. This example shows that as a point of principle boundary terms \emph{cannot} be dismissed when formulating predictions. In particular, the boundary part of the action \eqref{eq:zeta-action3}, that we have ignored here, will also contribute to the correlation functions (for general initial states).

Whether the above arguments point towards the need for a time-dependent EFT to be supplemented with boundary operators, or that symmetry requirements for bulk operators differ from those for boundary ones in EFT, is outside the scope of the present work. We do believe, however, that the above arguments do point towards such a possibility (also see, e.g., \cite{Collins:2012nq,Collins:2013kqa}).


\section{Summary}
\label{sec:summary}

\begin{figure}[htpb]
\begin{center}
\includegraphics[width=11cm]{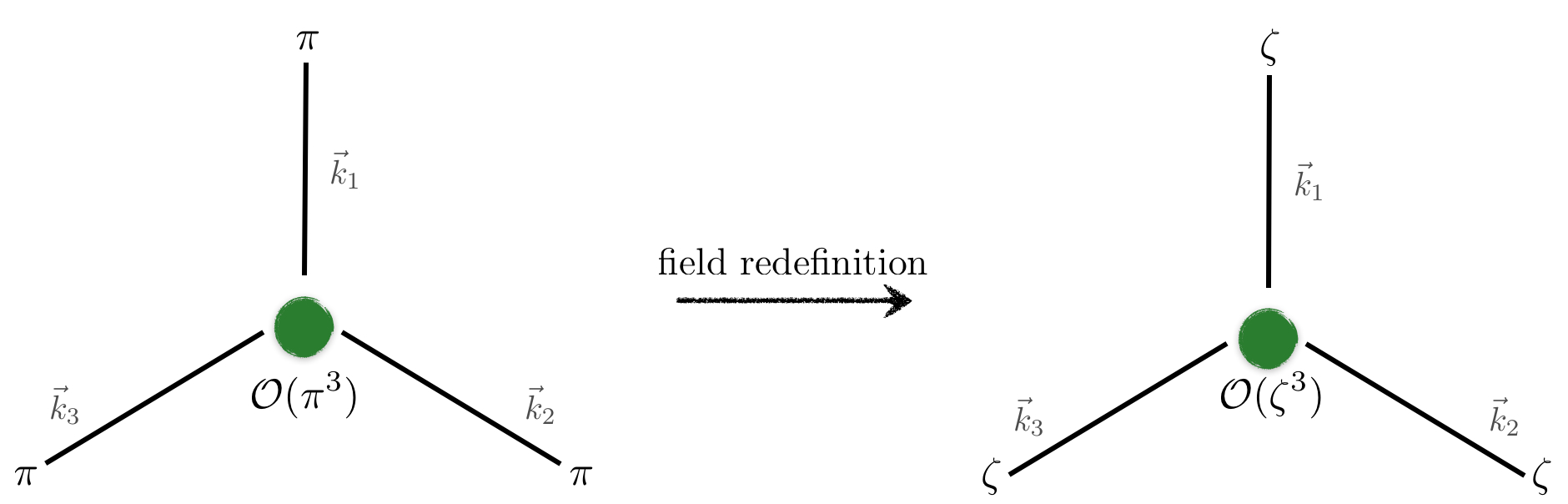}
\caption{The three-point correlators in the Goldstone picture are related to those in the primordial perturbation framework through a field redefinition, which is necessary for comparison with observations.}
\label{fig:3ptredefinition}
\end{center}
\end{figure}

Extracting observable quantities for inflation out of correlations of primordial perturbations is one of the key aspects of early Universe cosmology. Our objective in this work was to reconcile the two known ways of obtaining these correlation functions, using the Horndeski and EFT actions described earlier. Whichever method one chooses, the physical correlators are the ones built from the primordial perturbation itself, as depicted in figure \ref{fig:3ptredefinition}, since $\zeta$ is the quantity which is conserved on super-horizon scales.

For $P(X,\phi)$ theories, we have shown that we can derive the cubic EFT action in its simplest form from the Horndeski operators by performing a linear field redefinition. We find the same operators as in the standard form of EFT, which are therefore compatible with the remaining symmetries during slow-roll inflation, but with different interaction coefficients. We find that to fully revert to the usual action, one needs to add a boundary term which becomes a placeholder for an important, extra contribution to the bispectrum of primordial fluctuations. This boundary term has the same effect as a convolution in momentum space had we performed a non-linear field redefinition \cite{Cheung:2007sv}. We hence conclude that even though our results agree with those in the literature, they reveal a more straightforward algorithm to obtain $\zeta$ correlators from the EFT action. This action is given in eq.\ \eqref{eq:S3pibd} discarding the boundary term, and it captures all relevant interactions for inflationary perturbations that started in a Bunch--Davies initial state of the interacting theory. 

For more general initial states, however, operators defined at the boundary which result from the change of gauge do contribute to the three-point correlation. Moreover, since the Horndeski action itself includes boundary terms which are not slow-roll suppressed, this implies that there is a set of boundary terms which will contribute, in general, to the bispectrum of perturbations. For $P(X,\phi)$ models these boundary terms can be found in \cite{Arroja:2011yj, Burrage:2011hd}. Consequently, if exploring the bottom-up approach of quantum field theory by which one is able to reconstruct the Lagrangian parameters from observables, great care needs to be taken, especially if one intends to explore the impact of general excited states on observations.

Having made the dictionary between the Horndeski and EFT methods more precise, it would be important to understand whether there exist any degeneracies between different templates that are used when interpreting CMB measurements of the bispectrum in these two pictures. Additionally, since boundary terms can be important for general initial states, and since it is not clear that the same boundary terms appear in different approaches to inflationary perturbations, this dictionary would be important when using observational data to obtain an insight on the initial pre-inflationary state. If a signal consistent with a non-Bunch--Davies initial state is found, the deviation from the simplest choice of vacuum could reveal interesting aspects of the very high energy regimes, perhaps even a mechanism by which the initial fluctuations were produced. On the contrary, if such a departure from Bunch--Davies is not compatible with the data, we still need to explain why the simplest choice of initial state works.

To conclude, our answer to the question posed by the title of this paper is that the EFT approach really is equivalent to what would have been obtained by working directly with the action in terms of $\zeta$. While this may sound obvious, what we have seen from our calculations is that this equivalence has some highly non-trivial aspects to it and relies heavily on understanding the effect of boundary terms, which are usually neglected when writing down effective theories. These boundary terms are in particular crucial for a reliable prediction for general excited initial states. Beyond this equivalence, we have shown that we can simplify calculations in the EFT considerably by realizing that we need only use the linear relation between $\zeta$ and the Goldstone mode $\pi$, at least at the level of the bispectrum, by slightly changing the coefficients of the operators in the EFT approach. It would be interesting to understand how our calculation might be extended to higher $n$-point functions, though we expect the technical aspects to be considerably more involved. Moreover, technically, our analysis only applies to $P(X,\phi)$ models. Going beyond this class might be of considerable interest in categorizing all single-field models, but we consider such a study to be beyond the scope of this paper.


\acknowledgments 
It is a pleasure to thank Lasha Berezhiani, Matteo Fasiello, Austin Joyce, Leonardo Senatore, and especially Joyce Byun and Andrew Tolley for very helpful discussions. We also thank the anonymous referee for a thorough reading of our manuscript and for helpful comments. 
We thank the organizers of the KITP conference ``Observations and Theoretical Challenges in Primordial Cosmology," held at the University of California, Santa Barbara, for a very stimulating environment and where ideas leading to this work were discussed. R.\,H.\,R. would like to thank the kind hospitality of the McWilliams Center for Cosmology at Carnegie Mellon University, where this work was completed. N.\,A. is supported by the McWilliams Fellowship of the Bruce and Astrid McWilliams Center for Cosmology. R.\,H. is supported in part by the Department of Energy under grant DE-FG03-91-ER40682. N.\,A. and R.\,H. are also supported by the New Frontiers in Astronomy and Cosmology program at the John Templeton Foundation.


\bibliographystyle{JHEPmodplain}
\bibliography{references}

\end{document}